\documentclass[12pt,preprint]{aastex}

\newcommand{\etal}{et al.}

\shorttitle{ULIRGs, AGN, \& Mergers}
\shortauthors{Draper \&  Ballantyne}

\begin{document}

\title{The Merger-Triggered Active Galactic Nuclei Contribution to the Ultraluminous Infrared Galaxy Population}


\author{A. R. Draper and D. R. Ballantyne}
\affil{Center for Relativistic Astrophysics, School of Physics,
  Georgia Institute of Technology, Atlanta, GA 30332}
\email{aden.draper@physics.gatech.edu}

\begin{abstract}

It has long been thought that there is a connection between ultraluminous infrared galaxies (ULIRGs), 
quasars, and major mergers.  Indeed, simulations show that major 
mergers are capable of triggering massive starbursts and quasars.  
However, observations 
by the {\em Herschel Space Observatory} suggest that, at least at high redshift, there may not always be a 
simple causal connection between ULIRGs and mergers.  Here, we combine 
an evolving merger-triggered AGN luminosity function with a merger-triggered starburst model 
to calculate the maximum contribution of major mergers to the ULIRG population.  We find that major mergers can account for the 
entire local population of ULIRGs hosting AGN and $\sim$25$\%$ of the total local ULIRG luminosity density.  
By $z\sim1$, major mergers can no longer account for the luminosity density of ULIRGs hosting AGN and 
contribute $\lesssim$12$\%$ of the total ULIRG luminosity density.  
This drop is 
likely due to high redshift galaxies being more gas rich and therefore able to achieve high star formation rates 
through secular evolution.  Additionally, we find that major mergers can account for the local population of warm 
ULIRGs.  This suggests that selecting high 
redshift warm ULIRGs will allow for the identification of high redshift merger-triggered 
ULIRGs.  As major mergers are likely to trigger very highly obscured AGN, a significant 
fraction of the high redshift warm ULIRG population may host Compton thick AGN.

\end{abstract}



\section{Introduction}
\label{sect:intro}

In the 1980s astronomers discovered a new class of infrared selected galaxies known as ultraluminous 
infrared galaxies (ULIRGs) and characterized by $L_{IR}>10^{12}$ L$_{\odot}$, where $L_{IR}$ is the 8--1000 
$\mu$m luminosity \citep[e.g.,][]{H84,S84a}.  Another important topic during this time period was 
the study of the evolution of quasars \citep[e.g.,][]{SG83}, a class of active galactic nuclei 
(AGN) where accretion onto the supermassive black hole at the center of a massive galaxy gives rise to 
$L_X>10^{44}$ erg s$^{-1}$, where $L_X$ is the 2--10 keV luminosity.  ULIRGs and quasars have similar bolometric 
luminosities (10$^{45}$--10$^{46}$ erg s$^{-1}$) and optical observations suggest many ULIRGs have nuclear sources of non-thermal 
ionizing radiation and disturbed morphologies \citep{S88}.  Thus, \citet{S88} suggested 
that when two gas rich galaxies merge, gas and dust will fall into the 
nucleus of the resulting galaxy, triggering a massive starburst and a quasar.  

Thirty years later, the connection between ULIRGs, AGN, and major mergers is still an area 
of active research.  Recent simulations show that, indeed, gas rich major mergers are capable of triggering 
large starbursts and bright AGN \citep[e.g.,][]{H06, Y09, H10, N10}. 
By looking for ULIRGs with strong X-ray emission or a power-law spectra in the {\em Spitzer Space Telescope} 
IRAC bands, studies have shown that AGN are common in ULIRGs \citep[e.g.,][]{H06, A07, Y09, D10, H10, N10}.
Furthermore, 
both the fraction of ULIRGs that host AGN and the fraction of ULIRGs whose bolometric luminosities are dominated by AGN emission, appear to increase strongly with luminosity \citep[e.g.,][]{V02, P05, G05, B06, D10, Nard10}.  Morphological 
studies have shown that a significant fraction of ULIRGs have disturbed morphologies, suggesting the galaxy 
has recently undergone a merger or interaction \citep[e.g.][]{D10, K10, Nard10}.  Additionally, \citet{V02} showed 
that the fraction of ULIRGs triggered by major mergers increases with luminosity.  Thus, it is expected that a significant 
fraction of ULIRGs host AGN and were triggered by gas rich major mergers.  However, these studies tend to focus on ULIRGs with $z\lesssim1$.

Far-infrared observations by the {\em Herschel Space Telescope} have opened a new window on the $z\gtrsim1$ ULIRG population.  Interestingly, 
{\em Herschel} observations show that at high redshift major mergers are not necessary to trigger ULIRGs 
\citep{S10}.  Analyzing {\em Herschel} observations of the Bo\"otes field, \citet{M12} 
point out that $\lesssim$30$\%$ of optically-faint $z\sim2$ ULIRGs show obvious signs of a recent merger.  Deep 
{\em Herschel} observations of the Great Observatories Origins Deep Survey (GOODS) and the Cosmological Evolution Survey (COSMOS) fields find that most $z\gtrsim1$ ULIRGs are not in a 
starburst mode of star formation \citep{E11,R11}.  Instead, the 
increased gas fraction in high redshift galaxies allows normal secular star formation to power ULIRGs \citep[e.g.,][]{D08,E11,R11,M12}.  
Recent cosmological simulations confirm these observational results, finding that more than half of high redshift ULIRGs can be accounted 
for through mechanisms other than major mergers \citep{N12}.  These results indicate that at high redshift the AGN-ULIRG connection 
may be quite different than the connection observed locally.


\citet{DB12} computed an AGN population model that constrains the space density and Eddington ratio evolution of 
AGN triggered by major mergers by considering the hard X-ray luminosity function (HXLF), X-ray AGN number counts, the 
X-ray background, and the local mass density of supermassive black holes.  Thus,  
a model of the major merger population can be computed by combining this description of the 
evolving luminosity function of merger-triggered AGN with the \citet{H10} model for the time evolution of 
merger-triggered starbursts.  Similarly, a model merger spectral energy distribution (SED) can be calculated by combining AGN 
infrared spectra computed with the photoionization code {\sc Cloudy} \citep{F98} with the \citet{R09} star formation templates.  
This method is used here to determine the maximum contribution of mergers to the ULIRG 
population at $z\lesssim1.5$.  A $\Lambda$CDM cosmology is assumed 
with $H_0=70$ km s$^{-1}$ Mpc$^{-1}$ and $\Omega_{\Lambda}=1.0-\Omega_M=0.7$.

\section{Calculations}
\label{sect:calc}

\subsection{AGN Model}
\label{sub:agn}

The evolving HXLF of major merger-triggered AGN presented by \citet{DB12} is used to determine the space density and 
luminosity distribution of merger-triggered AGN.  \citet{DB12} combined an observationally motivated major merger  
rate \citep{H10a,T10a} and a theoretical AGN light curve \citep{HH09} to determine the contribution of major 
merger-triggered AGN to the HXLF.  \citet{HH09} parametrize the AGN light curve as
\begin{equation}
\lambda(t)=\left[1+\left(\frac{|t-t_Q|}{t_Q}\right)^{1/2}\right]^{-2/\beta},
\label{eq:lightcurve}
\end{equation}
where $\lambda(t)$ is the AGN Eddington ratio at time $t$ years after the AGN was triggered, $t_Q=t_0 \eta^{\beta}/(2\beta\ln 10)$, 
and $t_0$, $\eta$, and $\beta$ are parameters describing the AGN lifetime, peak Eddington ratio, and light curve slope, respectively.  
\citet{DB12} find that the best fit to the observed HXLF, X-ray AGN number counts, X-ray background, and mass density of 
supermassive black holes is achieved when $t_0=2.5\times10^8$ yrs, $\eta$ = 2.5, and 
$\beta$ = 0.7 for AGN triggered by mergers.  It is assumed that all mergers trigger AGN and therefore this model provides an upper 
limit to the contribution of major mergers to the AGN population.

Here, this merger-triggered AGN HXLF is converted to an infrared luminosity function using AGN  
SEDs computed using {\sc Cloudy} version C08.00 \citep{F98} as described in Section 2 of \citet{DB11}.  These SEDs cover 
the sub-mm to very hard X-ray wavelength regimes and incorporate the direct emission from the AGN, the diffuse emission radiated 
along the line of sight by the obscuring material around the AGN, and the emission reflected off the inner surface of the obscuring 
cloud.  As in \citet{DB11}, the neutral hydrogen density of the clouds is assigned such that Compton thin clouds have $n_H=10^4$ 
cm$^{-3}$ and Compton thick (CT) clouds have $n_H=10^6$ cm$^{-3}$, in accordance with the observed densities of typical molecular 
clouds.  The inner radius of the obscuring material is assumed to be $\sim$10 pc.  As we
compare against observations of an ensemble of sources, instead of fitting individual sources, a simple AGN torus model is appropriate 
for this study.  Moreover, as discussed in Section \ref{sect:results}, the infrared emission of ULIRGs is dominated by star formation 
processes, thus the results presented here are not dependent on the torus model used.

\subsection{Starburst Model}
\label{sub:star}

The evolving star formation rate, $\dot{M}_*(t)$, of the merger-triggered starburst is determined using the model of \citet{H10}.  Thus, 
\begin{equation}
\frac{dt}{d\dot{M}_*}=t_*\ln(10)\exp{\left(\frac{-\dot{M}_*}{M_{sb}/t_*}\right)},
\label{eq:SFR}
\end{equation}
where $t_*$ = 0.1 Gyr is the timescale of the merger-triggered starburst and $M_{sb}$ is the total mass of stars born during the merger-triggered starburst.    
Once $\dot{M}_*$ is determined, $L_{IR}$ due to star formation is calculated using \citep{H10}\footnote{Using the \citet{K98} conversion factor, 5.8 
$\times10^{9}$ L$_{\odot}$/(M$_{\odot}$ yr$^{-1}$), yields an infrared luminosity density 
$\sim$10$\%$ lower.}
\begin{equation}
L_{IR}=1.1\times10^{10}\left[\frac{\dot{M}_*}{1 M_{\odot} yr^{-1}}\right]L_{\odot}. 
\label{eq:conv}
\end{equation}

In order to combine the AGN and starburst models, the time delay between the triggering of the starburst and the 
triggering of the AGN must be considered.  Observational studies and simulations show that $\lambda$ tends to 
peak $\sim$100 Myr after $\dot{M}_*$ peaks \citep{D07, S07, S09, RZ10, W10, H12}.  Therefore, the starburst 
is triggered such that the time delay between the peak star formation rate, $\dot{M}_*^{peak}$, and the peak AGN 
Eddington ratio, $\lambda^{peak}$, $\Delta t$ = 100 Myr.  Scenarios where $\Delta t=0$ and 1 Gyr are also considered.  Once 
the merger occurs, the AGN light curve of Equation \ref{eq:lightcurve} and the star formation rate evolution of Equation \ref{eq:SFR} 
determine the evolution of the system.

\begin{figure*}
\begin{center}
\includegraphics[angle=0,width=0.95\textwidth]{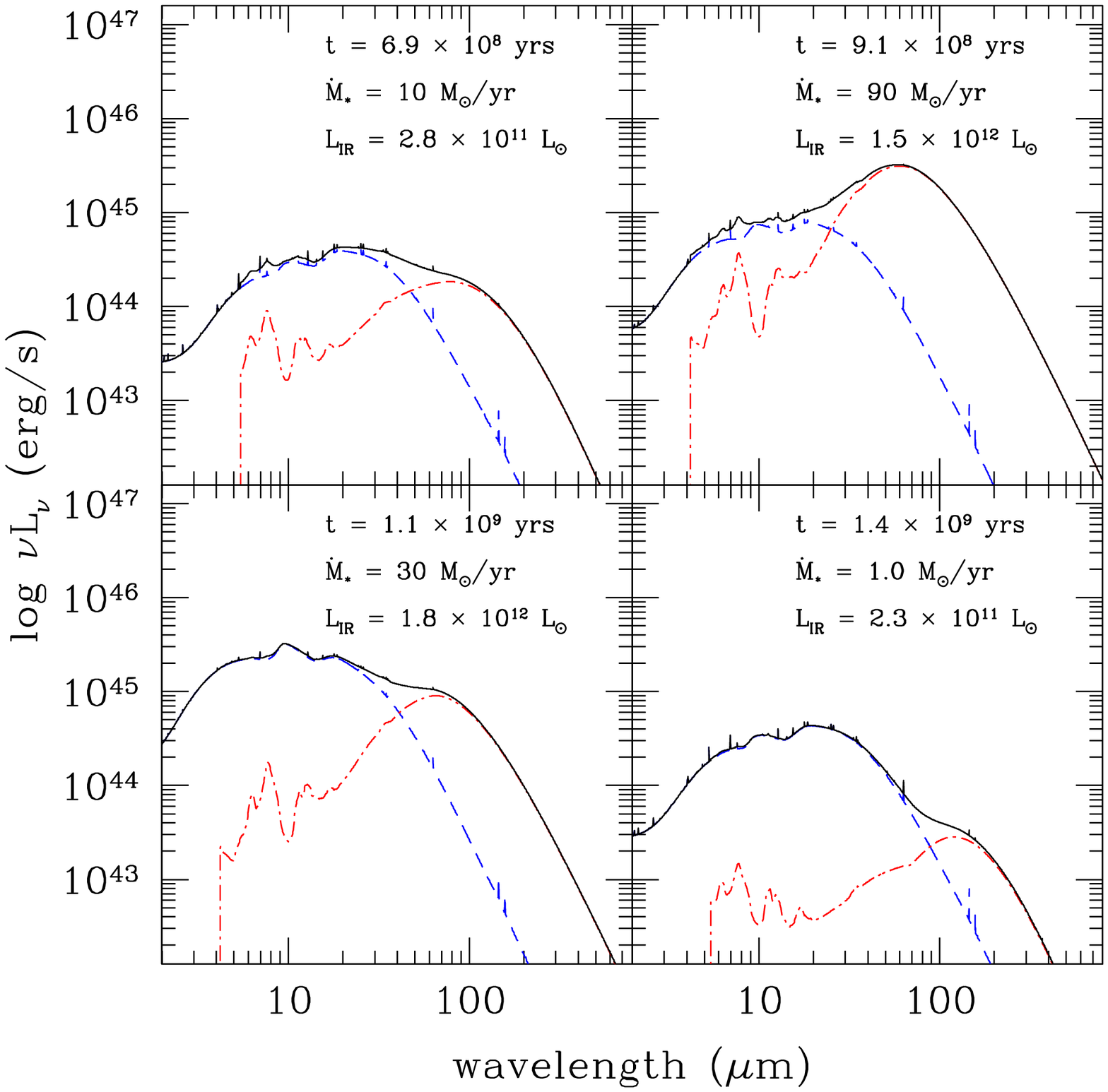}
\end{center}
\caption{Infrared SEDs of an AGN-starburst galaxy with black hole mass $M_{\bullet}$ = 10$^8$ M$_{\odot}$ and obscuring column density $N_H$ $\sim$ 10$^{23}$ cm$^{-2}$, at various times after being triggered by a major merger.  The dashed blues lines show the AGN infrared SEDs computed with {\sc Cloudy} and the red dot-dashed lines show the \citet{R09} starburst SEDs.  The solid black lines show the SED of the AGN-starburst galaxy.  In the upper left plot, the AGN and starburst both contribute approximately half of $L_{IR}$.  The upper right plot shows the SED at $\dot{M}_*$ $\approx$ $\dot{M}_*^{peak}$, and thus the starburst dominates $L_{IR}$.  The lower left plot shows the SED at $\lambda$ $\approx$ $\lambda^{peak}$.  The AGN dominates $L_{IR}$ in both of the bottom plots.}
\label{fig:sed}
\end{figure*}

Thus, the space density of major mergers at redshift $z$ is set by the merger rate.  
The starburst model is triggered at $z+\Delta z$, where $\Delta z$ is the appropriate change in redshift for $\Delta t$ at $z$.  
The corresponding AGN is triggered at $z$.  The evolution of the starburst is governed by Equation \ref{eq:SFR} and the AGN luminosity 
evolves according to Equation \ref{eq:lightcurve}.  The infrared luminosity of the merger remnant galaxy is then determined using 
the AGN SEDs calculated with {\sc Cloudy} and the \citet{R09} starburst SED templates.  The \citet{R09} templates are based on the SEDs of local, 
pure star forming galaxies and include poly-cyclic aromatic hydrocarbon (PAH) emission features.  Figure \ref{fig:sed} shows an example SED at 
four different times, $t$, after the major merger: before $\dot{M}_*$ reaches $\dot{M}_*^{peak}$ (upper left), $\dot{M}_*\approx\dot{M}_*^{peak}$ 
(upper right), $\lambda\approx\lambda^{peak}$ (lower left), and after $\lambda$ reaches $\lambda^{peak}$ (lower right). 
With the merger-triggered AGN HXLF, $d\Phi_{X}/d(\log L_X)$, and merger SEDs set, the merger infrared luminosity function, 
$d\Phi_{IR}/d(\log L_{IR})$, is calculated as 
\begin{equation}
\frac{d\Phi_{IR}(L_{IR},z)}{d(\log L_{IR})}=\frac{d\Phi_X(L_X,z)}{d(\log L_X)}\frac{d(\log L_X)}{d(\log L_{IR})},
\label{eq:phiIR}
\end{equation}
where $L_X$ and $L_{IR}$ are computed using the combined AGN-starburst SEDs.  The infrared luminosity density, $\Psi$, is then computed as
\begin{equation}
\Psi_{merger}(z)=\int_{L_{IR}^{min}}^{L_{IR}^{max}}L_{IR}\frac{d\Phi_{IR}(L_{IR}, z)}{d(\log L_{IR})}d(\log L_{IR}),
\label{eq:ld}
\end{equation}
where $L_{IR}^{min}=10^8$ L$_{\odot}$ and $L_{IR}^{max}=10^{13.5}$ L$_{\odot}$.  To calculate the ULIRG $\Psi_{merger}$, 
$L_{IR}^{min}$ is increased to 10$^{12}$ L$_{\odot}$.  The $\Psi_{merger}$ calculated here is then compared with the infrared luminosity density 
of AGN and their host galaxies, $\Psi_{AGN}$, measured by \citet{G10,G11} and the total infrared luminosity density, 
$\Psi_{total}$, measured by \citet{LF05}.

The major merger rate is set by parametrizing observations 
and simulations \citep{H10a, T10a} and the AGN light curve parameters are set by fitting observations of the 
AGN HXLF \citep{DB12}.  The evolution of the merger-triggered starburst model is parametrized 
by fitting results of simulations \citep{H10}.  This leaves only $M_{sb}$ and $\Delta t$ as free parameters.  
We explore $9.0\le\log M_{sb}$/M$_{\odot}\le11$ in steps of 0.25, assuming all mergers result in a similar value 
of $M_{sb}$.  The values $\Delta t$ = 0, 100 
Myr, and 1 Gyr are considered.  The maximum ULIRG $\Psi_{merger}$/$\Psi_{AGN}$ and $\Psi_{merger}$/$\Psi_{total}$ are then investigated.


%



\section{Results}
\label{sect:results}

\begin{figure*}
\begin{center}
\includegraphics[angle=0,width=0.95\textwidth]{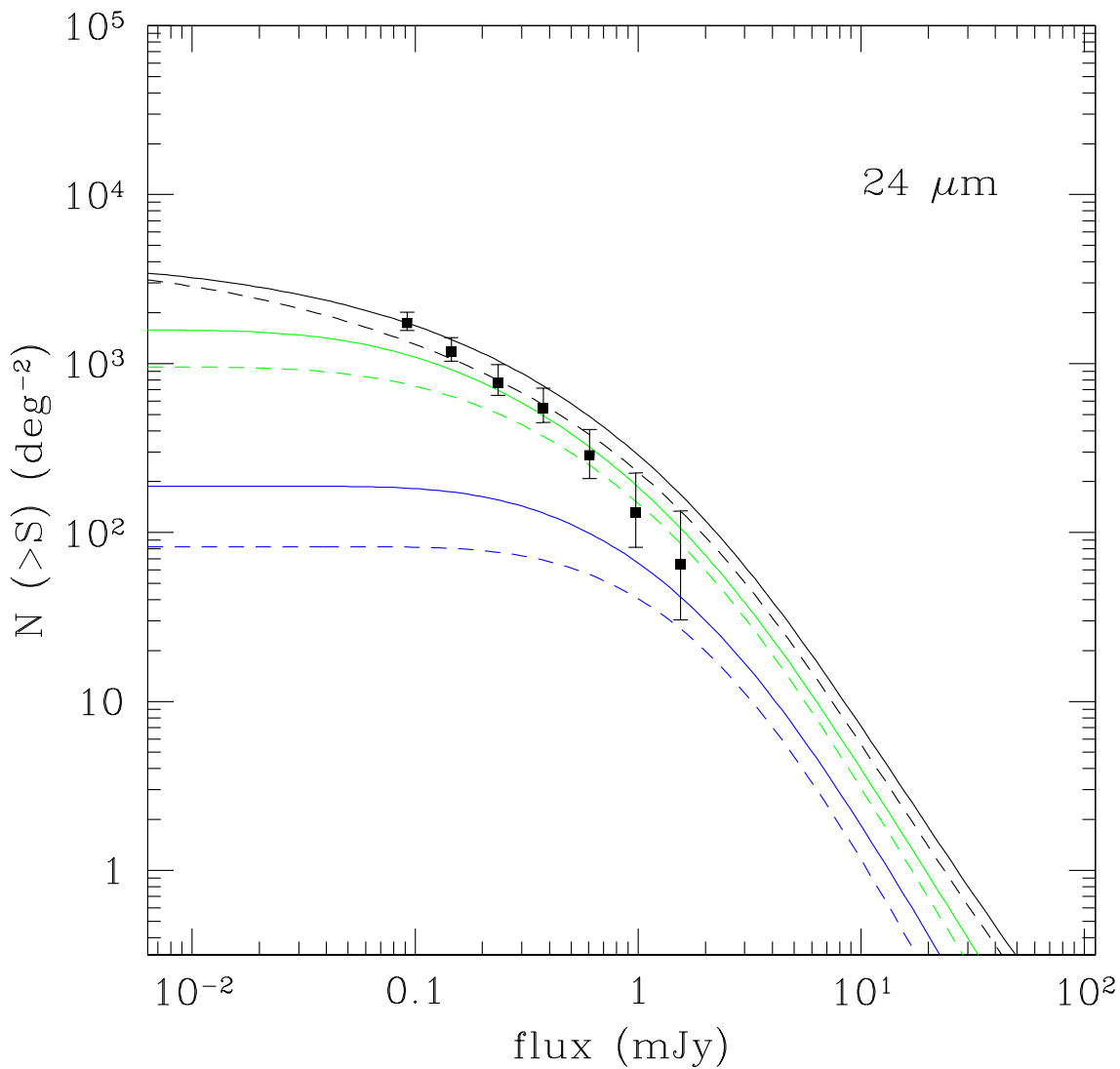}
\end{center}
\caption{24 $\mu$m number counts for mergers with $\Delta t$ = 100 Myr, $M_{sb}$ = 10$^{10}$ M$_{\odot}$, and $L_X>10^{42}$ erg s$^{-1}$.  The blue lines show the ULIRG number count and the green lines show the LIRG ($L_{IR}$ $>$ 10$^{11}$ L$_{\odot}$) number count.  The black lines show the 24 $\mu$m number count of all mergers.  The solid lines show the number counts for major mergers and the dashed lines show the number counts for the merger-triggered AGN only.  Data points show the number count of X-ray selected AGN in the GOODS field \citep{T06}.}
\label{fig:numcts}
\end{figure*}

In order to prevent over-predicting the maximum $\Psi_{merger}$, we consider the 24 $\mu$m number count of 
X-ray selected AGN following Equation 1 of \citet{DB11}.  If $M_{sb}\gtrsim10^{10.25}$ M$_{\odot}$ 
the bright end of the 24 $\mu$m number count is over-predicted by a factor $\gtrsim$2.  These same models 
also over-predict the local ULIRG
$\Psi_{total}$.  Thus, 10$^{10.25}$ M$_{\odot}$ is an upper limit on the average $M_{sb}$ for the 
population of major mergers.  Figure \ref{fig:numcts} 
shows the number count for mergers with $\Delta t$ = 100 Myr and $M_{sb}=10^{10}$ M$_{\odot}$.  
As the observed number count is from the GOODS fields \citep{T06}, it is not surprising that the model slightly over-predicts 
the bright end of the observed count since GOODS is a narrow field survey and likely misses bright, rare sources which are 
better accounted for by wide field surveys.



\begin{figure*}
\begin{center}
\includegraphics[angle=0,width=0.95\textwidth]{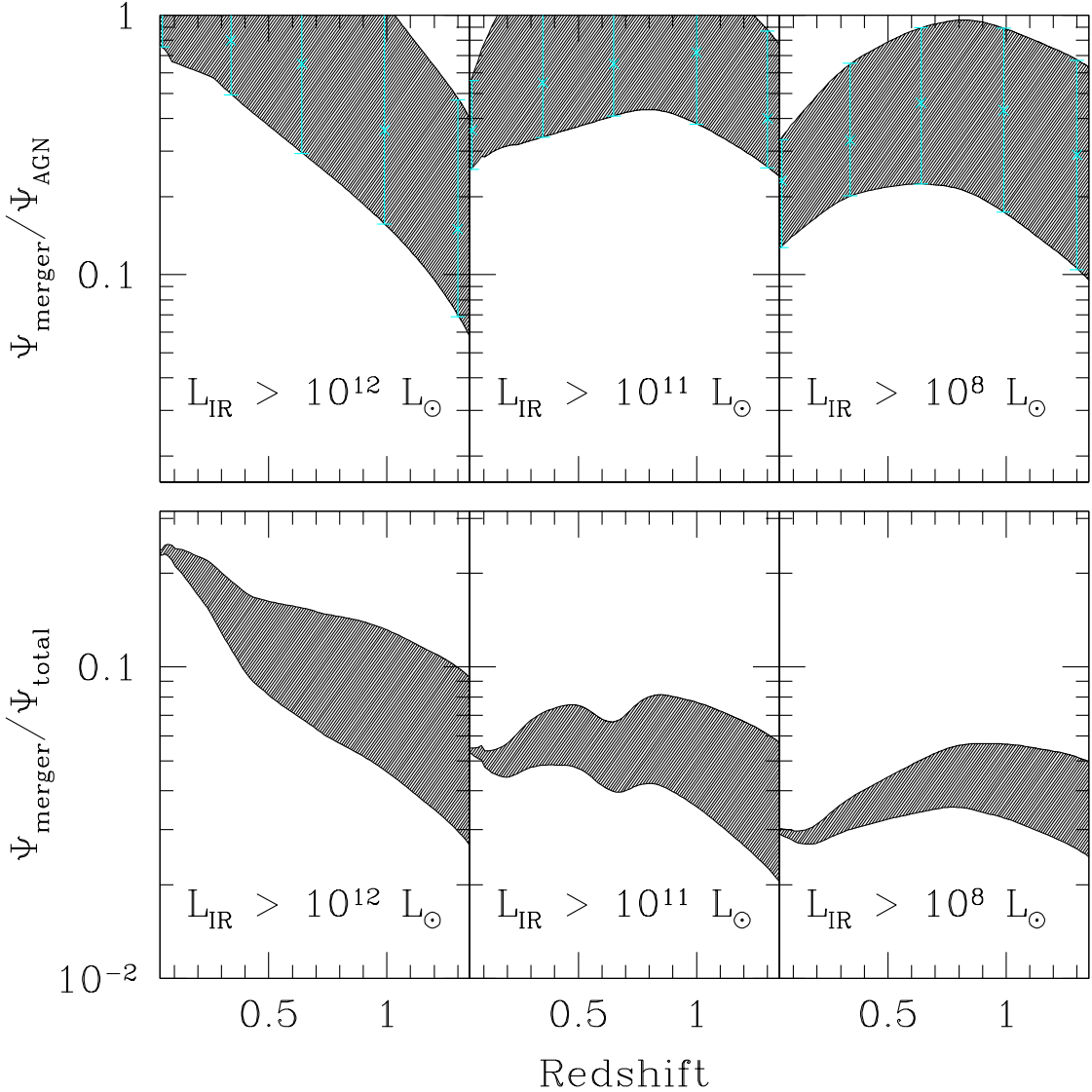}
\end{center}
\caption{Fraction of $\Psi_{AGN}$ and $\Psi_{total}$ contributed by major mergers with $\Delta t$ = 100 Myr and $M_{sb}$ = 10$^{10}$ M$_{\odot}$.  The top row shows the fraction $\Psi_{AGN}$ that can be accounted for by mergers for ULIRGs, LIRGs, and sources with $L_{IR}$ $>$ 10$^8$ L$_{\odot}$.  The observed $\Psi_{AGN}$ is determined by fitting the data points from \citet{G10,G11}, which are shown as the cyan points.  The bottom row shows the fraction of $\Psi_{total}$, as reported by \citet{LF05}, that can be accounted for by major mergers for ULIRGs, LIRGs, and sources with $L_{IR}$ $>$ 10$^8$ L$_{\odot}$.}
\label{fig:frac}
\end{figure*}

Figure \ref{fig:frac} shows the maximum ULIRG $\Psi_{merger}$/$\Psi_{AGN}$ and ULIRG $\Psi_{merger}$/$\Psi_{total}$  
that does not significantly over-predicting the 24 $\mu$m number count.  For the model 
shown in Figure \ref{fig:frac}, $\Delta t$ = 100 Myr and $M_{sb}=10^{10}$ M$_{\odot}$.  
The top left frame of Figure \ref{fig:frac} shows that mergers can account for the local ULIRG $\Psi_{AGN}$, however, 
as shown in the lower left frame, $\Psi_{merger}$/$\Psi_{total}\lesssim0.26$ for local ULIRGs.  
Interestingly, if we remove the starburst from our SEDS, we find that emission from AGN alone can contribute $\lesssim$20$\%$ of the local ULIRG $\Psi_{AGN}$. 
Thus, we confirm that the $L_{IR}$ of ULIRGs hosting AGN tends to be dominated by star formation.  

If $M_{sb}=10^{10}$ M$_{\odot}$, major mergers can account for the local ULIRG $\Psi_{AGN}$ for $\Delta t$ = 0 and 1 Gyr. Thus, 
the ability for mergers to account for the local AGN ULIRG population is not strongly dependent on $\Delta t$.  The local ULIRG $\Psi_{merger}$/$\Psi_{total}$  reduces to $\sim$0.20 for both $\Delta t$ = 0 and 1 Gyr.
If $\Delta t$ = 100 Myr and $M_{sb}\lesssim10^{9.75}$ M$_{\odot}$, mergers cannot account for the local 
population of ULIRGs hosting AGN.  Observations and simulations suggest that $M_{sb}$ is proportional 
to the galaxy stellar mass, $M_*$, such that $M_{sb}(M_*)=fM_*$ with the fraction $f$ on the order of, 
but $<$0.1 \citep{H10,Z12}.  If we assume a distribution of $M_{sb}(M_*)$ defined by $M_{sb}(M_*)=0.1M_*$, 
with the distribution of $M_*$ described by the \citet{PG08} stellar mass function, the AGN number count and 
the local ULIRG $\Psi_{total}$ are over-predicted by at least a factor of 2.  However, if $f=0.05$, the model 
predictions are in agreement with the AGN number count and the local ULIRG $\Psi_{merger}$/$\Psi_{total}=0.28$,
only 0.02 higher than the simple calculation with a single $M_{sb}$.  If $f=0.05$, 
$\langle M_{sb}(M_*)\rangle=10^{10.05}$ M$_{\odot}$.  Therefore, for $\Delta t$ = 0--1 Gyr and $M_{sb}\sim10^{10}$ M$_{\odot}$, the 
local ULIRG $\Psi_{merger}$/$\Psi_{AGN}\approx1.0$ and the local ULIRG $\Psi_{merger}$/$\Psi_{total}\approx$ 0.20--0.28.

\section{Discussion}
\label{sect:disc}

Major mergers can account for at most a quarter of the local ULIRG $\Psi_{total}$.  
This suggests that a large fraction of local ULIRGs are triggered by mechanisms other than the coalescence of two massive gas 
rich galaxies, such as minor mergers, interactions, and secular processes.  By $z\sim1$, the ULIRG $\Psi_{merger}$/$\Psi_{AGN}<1$ and by 
$z\sim1.25$, the ULIRG $\Psi_{merger}$/$\Psi_{AGN}\lesssim0.5$.  
At $z\sim1$, the ULIRG $\Psi_{merger}$/$\Psi_{total}\lesssim0.12$.  Indeed, simulations by \citet{H10} predict 
that at $z=1$, the ULIRG $\Psi_{merger}$/$\Psi_{total}\approx0.08$, in good agreement with the findings of this study.  
As major mergers are much more common at $z>1$ 
than $z<1$, this suggests that secular processes are even more important for triggering ULIRGs at high redshift.  {\em Herschel} has 
provided observational evidence that major mergers are not necessary at $z\gtrsim1$ 
to trigger ULIRGs \citep{S10}; a finding that has been confirmed by simulations \citep{N12}.

At $z\gtrsim1$ the majority of ULIRGs appear to be scaled up versions of local normal star forming galaxies.  These 
high redshift ULIRGs tend to have cooler far-infrared dust temperatures \citep{M12} and stronger PAH emission \citep{E11} 
than local ULIRGs.  Furthermore, by comparing the 8 $\mu$m flux to the total infrared flux of high redshift ULIRGs, 
\citet{E11} find that most $z\gtrsim1$ ULIRGs lie on the infrared main sequence and are in a normal star forming mode 
of evolution.   Similarly, \citet{B12} and \citet{K12} find that at $z\sim0.7$ and $z\sim2$, AGN tend to be hosted 
by galaxies on the main sequence of star formation \citep[see][]{E11}.  Observations and simulations both 
point out that high redshift galaxies are more gas rich than local galaxies 
\citep[e.g.,][]{T10a, Di12, N12}. It is likely that galaxies with large reservoirs of gas and dust are capable 
of fueling ULIRGs without being triggered by a major merger.

\citet{E11} do find that some high redshift ULIRGs are in a starburst mode, possibly triggered by a merger.  Moreover, \citet{E11} 
find evidence for a population of very highly obscured AGN embedded in these compact dusty starbursts.  Other observational studies 
also find CT AGN ($N_H>10^{24}$ cm$^{-2}$) candidates in high redshift dusty starburst galaxies \citep[e.g.,][]{D10,M12}.  
\citet{F99} explains that the gas and dust funneled into the central regions of the merger remnant galaxy will fuel a burst of star formation, 
rapid black hole accretion, and will obscure the resulting AGN.  Thus, a significant fraction of CT AGN are expected to be recently 
triggered, likely by a major merger, and accreting very rapidly \citep[e.g.,][]{DB10}.  The results of this study are consistent with the merger-triggered CT AGN scenario, 
but, as $\Psi_{merger}$/$\Psi_{total}\lesssim0.12$ for ULIRGs at $z\sim1$, the contribution of these merger-triggered CT AGN to the $z\gtrsim1$ ULIRG population must be fairly small.  The 
fraction of AGN that are CT is hard to observationally constrain due to the severe obscuration that defines CT AGN and 
AGN population models predict a wide range for the CT fraction \citep[see][]{B11}.  Increasing the fraction of AGN that are CT in this model by a factor of 1.5 increases the $z\sim1$ ULIRG $\Psi_{merger}$/$\Psi_{total}$ by $\sim$0.01.

Local ULIRG SEDs are often divided into two groups, warm and cool, where warm ULIRG SEDs are characterized by 
$f_{25\mu m}/f_{60\mu m}>0.2$ \citep[e.g.,][]{AH06}, where $f_{25\mu m}$ is the 25 $\mu$m flux and $f_{60\mu m}$ is the 60 $\mu$m flux.
According to \citet{AH06}, warm ULIRGs are 15--30$\%$ of the Bright Galaxy Survey sources \citep{S88b}.  
The merger model used here does produce warm ULIRG SEDs with $f_{25\mu m}/f_{60\mu m}\gtrsim0.5$ at all redshifts; 
thus mergers can account for the local population of warm ULIRGs. \citet{E11} found that galaxies on the main sequence of 
normal star formation, including galaxies hosting AGN, tend to have cooler dust temperatures than star-bursting galaxies.   
Because 
starbursts triggered by major mergers are expected to be more compact than secular star formation, merger-triggered starbursts will be 
characterized by higher dust temperatures than normal star formation \citep[e.g.,][]{E11, M12}.  Indeed, observations of local ULIRGs 
show that compact ULIRGs are more likely to host an AGN than less compact ULIRGs \citep{Nard10}.  Analyzing simulations of major mergers, \citet{Y09} 
find that warm ULIRGs are likely to be galaxies evolving from the star formation dominated merger phase to the AGN-starburst post-merger phase.
Thus, by combining {\em Herschel} and {\em Spitzer} or {\em Wide-field Infrared Survey Explorer} (WISE) observations to select warm ULIRGs, 
the population of ULIRGs hosting 
merger-triggered AGN can be identified.  
As discussed above, the population of ULIRGs hosting AGN triggered by major 
mergers is likely to include a significant fraction of CT AGN.  Thus seeking out high redshift ULIRGs with warm SEDs will also lead to the 
identification of high redshift CT AGN.


By combining the evolving luminosity function of AGN triggered by major mergers calculated by \cite{DB12} with the merger-triggered 
starburst model of \citet{H10}, we computed an upper limit for the major merger contribution to the ULIRG population.  Locally, major 
mergers can account for the observed population of ULIRGs hosting AGN and ULIRGs with warm SEDs, but the local ULIRG $\Psi_{merger}$/$\Psi_{total}\lesssim0.26$.  
By $z\sim1$, major merger-triggered ULIRGs hosting AGN can no longer account for the population of ULIRGs 
observed to have AGN signatures.  Indeed, the ULIRG $\Psi_{merger}$/$\Psi_{AGN}\lesssim0.50$ by $z\sim1.25$.  Furthermore, 
at $z\sim1$, the ULIRG $\Psi_{merger}$/$\Psi_{total}\lesssim0.12$.  
Combining observations by {\em Herschel} and {\em Spitzer} or WISE to identify high 
redshift ULIRGs with warm SEDs is a good tool for identifying the population of ULIRGs hosting merger-triggered AGN, a large 
fraction of which are expected to be CT.









\acknowledgments
The authors thank the referee for helpful comments which improved this letter.  This work was supported by NSF award AST 1008067.


{}



%
%

\end{document}